\documentclass{PoS}

\title{Applying Mellin-Barnes technique and Gr\"obner bases to the three-loop
static potential}

\ShortTitle{Applying Mellin-Barnes technique and Gr\"obner bases}

\author{Alexander V. Smirnov\\
Scientific Research Computing Center of Moscow State University, Russia\\
E-mail: \email{asmirnov@rdm.ru}}

\author{\speaker{Vladimir A. Smirnov}%
\\
Nuclear Physics Institute of Moscow State University, Russia\\
E-mail: \email{smirnov@theory.sinp.msu.ru}}

\author{Matthias Steinhauser\\
Institut f{\"u}r Theoretische Teilchenphysik,
Universit{\"a}t Karlsruhe, Germany\\
E-mail: \email{matthias.steinhauser@uka.de}}

\abstract{
The Mellin--Barnes technique to evaluate master integrals and the algorithm
called {\tt FIRE} to solve IBP relations with the help of Gr\"obner bases are
briefly reviewed. In {\tt FIRE}, an extension of the classical Buchberger
algorithm to construct Gr\"obner bases is combined with the well-known Laporta
algorithm. It is explained how both techniques are used when evaluating the
three-loop correction to the static QCD quark potential. First results are
presented: the coefficients of $n_l^3$ and $n_l^2$, where $n_l$ is the number of
light quarks.}

\FullConference{8th International Symposium on Radiative Corrections \\
                 October 1-5, 2007\\
                 Florence, Italy}

\newcommand{\ep}{\varepsilon}

\newcommand{\be}{\begin{equation}}
\newcommand{\ee}{\end{equation}}
\newcommand{\bea}{\begin{eqnarray}}
\newcommand{\eea}{\end{eqnarray}}

\newcommand{\Gm}{\Gamma}

\newcommand{\lm}{\lambda}

\newcommand{\dd}{\mbox{d}}

\newcommand{\nn}{\nonumber}

\begin{document}


\section{Introduction}

At the modern level of analytic calculations in elementary particle physics one
needs to evaluate thousands and millions of multiloop Feynman integrals.
To evaluate a family of Feynman integrals which have the same structure
of the integrand and differ by powers of propagators (indices) the standard
strategy is to apply integration-by-parts (IBP) relations and solve the
problem in two steps: a reduction of any given Feynman integral to so-called
master integrals and the evaluation of the master integrals. In this
contribution we describe the computation of Feynman integrals needed for the
evaluation of the three-loop corrections to the static QCD quark potential. In
particular we describe the use of the Mellin--Barnes technique to evaluate master
integrals and the application of Gr\"obner bases to solve the IBP relations.

In Section~2 we present a brief review of the method of
Mellin--Barnes (MB) representation
and exemplify it by the evaluation of a non-trivial integral
contributing to the three-loop potential.
Afterwards we explain in Section~3 the main features of the algorithm called
{\tt FIRE} (Feynman Integral REduction)
which is based on an extension of the classical Buchberger
algorithm to construct Gr\"obner bases (see, e.g., Ref.~\cite{Buch}). In {\tt
  FIRE} Gr\"obner bases are naturally combined with the well-known Laporta
algorithm.
Finally, in Section~4 we present the first results of this evaluation: the
contributions proportional to $n_l^3$ and $n_l^2$, where $n_l$ is the number of
massless quarks.


\section{Mellin--Barnes technique}

\begin{figure}[b]
  \begin{center}
\begin{tabular}{ccc}
\begin{minipage}{14em}
    \includegraphics[width=\textwidth]{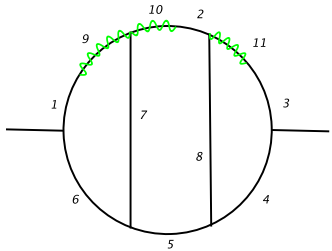}
\end{minipage}
    & \hspace*{2em} &
\begin{minipage}{20em}
    \caption{\label{fig::350}Feynman integral appearing in the calculation of
      the three-loop corrections to the static potential. The straight and
      wiggled lines correspond to massless scalar and static propagators,
      respectively. The numbers next to the lines refers to the corresponding
      index $a_i$.}
\end{minipage}
\end{tabular}
  \end{center}
\end{figure}

The MB representation
\bea
\frac{1}{(X+Y)^{\lm}} = 
\int_{- i  \infty}^{+ i  \infty}
\frac{Y^z}{X^{\lm+z}} \frac{\Gm(\lm+z) \Gm(-z)}{\Gm(\lm)}
\frac{\dd z}{2\pi  i}
\label{MB}
\eea
can be applied to replace a sum of two terms raised
to some power by their products in some powers.
For planar diagrams, experience shows that a minimal number of
MB integrations is achieved if one introduces them loop by loop,
i.e. one derives a MB representation for a one-loop subintegral,
inserts it into a higher two-loop integral, etc.
Consider, for example, the dimensionally regularized Feynman integral
of Fig.~\ref{fig::350} which we denote by $F(a_1,\ldots,a_{11})$.
A straightforward implementation of the loop-by-loop strategy
leads to a six-fold MB representation which reads
\bea
F(a_1,\ldots,a_{11})
= \frac{\left(i\pi^{d/2} \right)^3 (-q^2)^{6-a_{1,\ldots,8}
-a_{9,10,11}/2-3\ep}2^{a_{9,11}-2}}{(v^2)^{a_{9,10,11}/2}
\sqrt{\pi} \prod_{i=1,3,4,6,7,8,9,11} \Gm(a_i)
\Gm(4  - a_{3,4,8,11} - 2 \ep) \Gm(4 - a_{1,6,7,9} - 2 \ep) }
\nn \\ &&  \hspace*{-145mm}\times
\frac{1}{(2\pi i)^6} \int_{-i\infty}^{+i\infty}
\prod_{j=1}^6 \left( \Gm(-z_j) \dd z_j\right)
\frac{\Gm(a_4 + z_{1,2}) \Gm(a_6 + z_{4,5}) \Gm(1/2 - z_3) \Gm(1/2 - z_6)}
{\Gm(a_5 - z_{2,5})
\Gm(a_{1,2,3,4,6,7,8} + a_{9,11}/2 + 2 \ep-4  + z_{1,\ldots,6})}
\nn \\ &&  \hspace*{-145mm} \times
\frac{\Gm(a_9/2 + z_6) \Gm(a_{11}/2 + z_3)
\Gm(2 - a_{3,4}- a_{11}/2  - \ep - z_1 + z_3)
\Gm(a_{11}/2-2 + a_{3,4,8} + \ep + z_{1,2,3})}
{\Gm(a_{10}/2+1/2 - z_{3,6})
\Gm(8 - a_{1,\ldots,8} -a_{9,10,11}/2- 4 \ep - z_{1,4} + z_{3,6})}
\nn \\ &&  \hspace*{-145mm} \times
\Gm(a_{1,\ldots,8} + a_{9,10,11}/2 -6+ 3 \ep +z_{1,4})
\Gm(6 - a_{1,2,3,4,6,7,8}- a_{9,10,11}/2 -3 \ep - z_{1,2,4,5})
\nn \\ &&  \hspace*{-145mm} \times
\Gm(2 - a_{4,8}-a_{11}/2  - \ep - z_{2,3})
\Gm(2 - a_{1,6} - a_9/2 - \ep - z_4 + z_6)
\Gm(2  - a_5- a_{10}/2 - \ep + z_{2,3,5,6})
\nn \\ &&  \hspace*{-145mm} \times
\Gm(2 - a_{6,7} - a_9/2 - \ep - z_{5,6})\Gm(a_{1,6,7} + a_9/2-2 + \ep + z_{4,5,6})
\,,
\eea
where $a_{3,4,8,11}=a_3+a_4+a_8+a_{11}$, $z_{1,2,3}=z_1+z_2+z_3$, etc. By
definition, any integration contour over $z_i$ should go to the right (left) of
poles of Gamma functions with $+z$-dependence ($-z$-dependence).

There are two strategies for resolving the singularities in $\ep$ in MB
integrals suggested in Refs.~\cite{MB1,MB2} (see also Chapter~4 of~\cite{books}).
The second one was formulated
algorithmically \cite{AnDa,Czakon}, and the corresponding public code
{\tt MB.m}~\cite{Czakon} has become by now a standard way to evaluate MB
integrals in an expansion in $\ep$. It can be combined with the program
{\tt AMBRE}~\cite{AMBRE} which can be used to derive MB representations in the
loop-by-loop approach. Using {\tt MB.m} and evaluating the resulting finite
MB integrals by corollaries of Barnes lemmas we have obtained, for example,
the following result for one of the master integrals occuring in the
reduction of $F(a_1,\ldots,a_{11})$
\be
F(1,\ldots,1,0,1)=-\frac{(i\pi^{d/2})^3}{(-q^2)^{3+3\ep} v^2}
\left[
\frac{56\pi^4}{135\ep}
+\frac{112 \pi^4}{135} + \frac{16 \pi^2\zeta(3)}{9}
+ \frac{8\zeta(5)}{3} +O(\ep)
\right]\,.
\ee

At the three- and four-loop level, the method of MB representation
was successfully applied in \cite{3b,BDS,Bern:2006ew,CSV}.

The loop-by-loop approach becomes problematic for non-planar diagrams.
A typical phenomenon is that factors such as $(-1)^{z_i}$  arise.
This means that the convergence of MB integrals at large values of Im$(z_i)$ is
no longer guaranteed. Moreover, poles in $\ep$ arise not only due to ``gluing''
of poles of different nature (a typical example is the product
$\Gm(\ep+z)\Gm(-z)$; there is no space between the first poles of the two gamma
functions if $\ep\to 0$) but also from the integration over large Im$(z_i)$.
A safe way to proceed with non-planar diagrams is to
start from an alpha (Feynman) parametric representation and apply
(\ref{MB}) to the two basic functions in this representation.
(See also Ref.~\cite{Czakon1} for a discussion of this problem.)


\section{{\tt FIRE}: Feynman Integral REduction}

There are three approaches to solve IBP relations~\cite{IBP} in a systematic way:
Laporta's algo\-rithm
\cite{Laporta}, Baikov's method \cite{Baikov} and
two approaches using Gr\"obner bases which are described in Refs.~\cite{Tar}
and~\cite{SS1,SS2,AS,SS3}, respectively.
To solve the reduction problems arising in the evaluation of the three-loop
potential we use the latter algorithm whose computer implementation is called
{\tt FIRE}. It is based on a generalized Buchberger
algorithm for constructing Gr\"obner-type bases associated with polynomials of
shift operators. This method was recently used to evaluate a family of nontrivial
three-loop Feynman integrals~\cite{AGG}.
Let us in the following describe some new features of this algorithm.

Similarly to Laporta's algorithm, in our approach we work in a given {\em
sector}, i.e. a domain of integer indices $a_i$ where some indices are positive
and the rest of the indices are non-positive. The aim is to express any
integral from the sector in terms of master integrals of this sector {\em
and} integrals from lower sectors, where at least one more index is
non-positive. It turns out that in the higher sectors (with a small number of
non-positive indices) the corresponding $s$-basis \cite{SS2,SS3,AS} (a kind
of a Gr\"obner basis) can be constructed easily (and, in most cases, even
automatically).

In the opposite situation where a lot of non-positive
indices occur, $s$-bases are constructed not so easily.
Usually there is the possibility to explicitly
perform an integration over some loop momentum
for general value of $\ep$ with results in terms of gamma
functions. A straightforward way to do this leads to multiple summations and
turns out to be impractical.
However, there is an  alternative approach which can be illustrated using
the example of diagram of Fig.~\ref{fig::350}:
consider the region $a_2,a_5,a_{10}\leq 0$ and $a_7,a_8>0$
(i.e. the union of the sectors with such restrictions).
In this situation one can integrate over the middle loop momentum $l$ which enters
the propagators of the central subgraph with five lines. By
constructing an $s$-basis for this region, it turns out possible
to solve the IBP relations for the corresponding subintegral over $l$ in
order to express any such subintegral in terms of master integrals.
In other words, the indices $a_2,a_5,a_{10},a_7,a_8$ can be reduced to
their boundary values, i.e.  $a_2,a_5,a_{10}=0,\;\;a_7,a_8=1$, up to integrals
that drop out from this region. Then, after using this reduction
procedure, it will be sufficient to use explicit integration
formulae only for the boundary values of the indices.
This replacement is very simple, without multiple summations.

It turned out possible to implement the solution of the
recursive problem for the subgraph in terms of the Feynman
integrals for the whole graph. In this reduction, pure powers
of the parameters which are external for the subgraph
transform naturally into the corresponding shift operators and their
inverse.
Integrals which are obtained from initial integrals by an explicit
integration over a loop momentum in terms of gamma function
usually involve a propagator with an analytic regularization by an amount
proportional to $\epsilon$ (and, sometimes, $2\epsilon$).
After this integration we obtain a two-loop reduction problem with seven
indices which is then solved by {\tt FIRE}.

After using Gr\"obner bases in higher sectors and an explicit integration in
lower sectors, it is still necessary to solve the reduction problem in a
relatively small number of intermediate sectors. In these cases we turn to
Laporta's algorithm implemented as part of {\tt FIRE}.


\section{Evaluating three-loop static quark potential}

The QCD potential between a static quark and its antiquark can be cast in the form
\begin{eqnarray}
  V(|{\vec q}|) &=& -{4\pi C_F\alpha_s\over{\vec q}^2}
  \Bigg[1+{\alpha_s\over 4\pi}a_1
  +\left({\alpha_s\over 4\pi}\right)^2a_2
  +\left({\alpha_s\over 4\pi}\right)^3
  \left(a_3+ 8\pi^2 C_A^3\ln{\mu^2\over{\vec q}^2}\right)
  +\cdots\Bigg]\,,
  \label{singlet}
\end{eqnarray}
where the renormalization scale of $\alpha_s$ is set to $\vec{q}^2$.
The one-loop contribution $a_1$ is known since almost 30 years and also the
two-loop term has already been computed end of the nineties. Furthermore
logarithmic contributions are known at three- and four-loop level.
Explicit results and the references are nicely summarized in
the review~\cite{Vairo:2007id}.
The non-logarithmic third-order term, $a_3$, is still unknown. It is
conveniently be parametrized in the form
\begin{eqnarray}
  a_3 &=& a_3^{(3)} n_l^3 + a_3^{(2)} n_l^2 + a_3^{(1)} n_l + a_3^{(0)}\,,
\end{eqnarray}
where $n_l$ denotes the number of massless quarks. Using the techniques
described above we evaluated the coefficients $a_3^{(3)}$ and the
$C_AT_F^2$ part of $a_3^{(2)}$ which read
\begin{eqnarray}
  a_3^{(3)} = - \left(\frac{20}{9}\right)^3 T_F^3\,,
  \quad
  a_3^{(2)}\Big|_{C_AT^2} =
  \left(\frac{12541}{243}
    + \frac{368\zeta(3)}{3}
    + \frac{64\pi^4}{135}
  \right) C_A T_F^2
  \,.
\end{eqnarray}


\noindent
{\bf Acknowledgements.}
This work was supported by RFBR, grant 05-02-17645, and by DFG through project
SFB/TR~9.




\begin{thebibliography}{99}


%
%

\bibitem{Buch}
B.~Buchberger and F.~Winkler (eds.),
{\em Gr\"obner Bases and Applications},
Cambridge University Press, 1998.

\bibitem{MB1}
V.A.~Smirnov, Phys. Lett.  B {\bf 460}  (1999) 397.

\bibitem{MB2}
J.B.~Tausk, Phys. Lett.  B {\bf 469}  (1999) 225.

\bibitem{books}
V.~A.~Smirnov, ``Evaluating Feynman Integrals,''
Springer Tracts Mod.\ Phys.\  {\bf 211} (2004) 1;
``Feynman integral calculus,''
{\it  Berlin, Germany: Springer (2006) 283 p}.

\bibitem{AnDa}
C.~Anastasiou and A.~Daleo,
JHEP {\bf 0610} (2006) 031.

\bibitem{Czakon}
M.~Czakon, Comput.\ Phys.\ Commun.\  {\bf 175} (2006) 559.

\bibitem{AMBRE}
J.~Gluza, K.~Kajda and T.~Riemann,
Comput.\ Phys.\ Commun.\  {\bf 177} (2007) 879;
J.~Gluza, F.~Haas, K.~Kajda and T.~Riemann,
arXiv:0707.3567.

\bibitem{3b}
V.A. Smirnov, Phys. Lett. B {\bf 547} (2002) 239;
Phys. Lett. B {\bf 567} (2003) 193.

\bibitem{BDS}
Z. Bern,  L.J. Dixon, and V.A. Smirnov,
Phys.\ Rev.\ D {\bf 72} (2005) 085001.

\bibitem{Bern:2006ew}
Z.~Bern, M.~Czakon, L.J.~Dixon, D.A.~Kosower and V.A.~Smirnov,
Phys.\ Rev.\  D {\bf 75} (2007) 085010.

\bibitem{CSV}
F.~Cachazo, M.~Spradlin and A.~Volovich,
JHEP {\bf 0607} (2006) 007;
Phys.\ Rev.\  D {\bf 75} (2007) 105011;
Phys.\ Rev.\  D {\bf 76} (2007) 106004.

\bibitem{Czakon1}
M.~Czakon, A.~Mitov and S.~Moch,
arXiv:0707.4139 [hep-ph].

\bibitem{IBP}
K.G.~Chetyrkin and F.V.~Tkachov, Nucl. Phys. B {\bf 192}  (1981) 159.

\bibitem{Laporta}
S.~Laporta, Int. J. Mod. Phys. A15 (2000) 5087;
T.~Gehrmann and E.~Remiddi, Nucl. Phys. B {\bf 601} (2001) 248, 287.

\bibitem{Baikov}
P.A.~Baikov, Phys. Lett.  B {\bf 385}  (1996) 404;
Nucl.~Instrum.~Methods A {\bf 389}  (1997) 347;
Phys.\ Lett.\  B {\bf 474} (2000) 385;
Nucl.\ Phys.\ Proc.\ Suppl.\  {\bf 116} (2003) 378;
Phys.\ Lett.\  B {\bf 634} (2006) 325;
V.A.~Smirnov and M.~Steinhauser, Nucl. Phys.  B {\bf 672} (2003) 199.

\bibitem{Tar}
O.V.~Tarasov, Acta  Phys. Polon. B {\bf 29} (1998) 2655;
Nucl. Instrum. Meth. A {\bf 534} (2004) 293.

\bibitem{SS1}
A.V.~Smirnov and V.A.~Smirnov,
JHEP {\bf 0601}, 001 (2006).

\bibitem{SS2}
A.V.~Smirnov and V.A.~Smirnov,
Nucl.\ Phys.\ Proc.\ Suppl.\  {\bf 160}, 80 (2006);
\\
V.A.~Smirnov,
Nucl.\ Phys.\ Proc.\ Suppl.\  {\bf 157}, 131 (2006).

\bibitem{AS}
A.V.~Smirnov,
JHEP {\bf 0604} (2006) 026.

\bibitem{SS3}
A.V.~Smirnov and V.A.~Smirnov,
PoS(ACAT) (2007) 085 [arXiv:0707.3993].

\bibitem{AGG}
A.G.~Grozin, A.V.~Smirnov and V.A.~Smirnov,
JHEP {\bf 0611}, 022 (2006).

\bibitem{Vairo:2007id}
  A.~Vairo,
  AIP Conf.\ Proc.\  {\bf 964} (2007) 102.


\end{thebibliography}
\end{document}